УДК 614.8

# ВЫБОР ЛУЧШЕГО РЕСУРСА МЕТОДОМ МАМДАНИ

Е.В. ЛЕГЧЕКОВА[1], кандидат физико-математических наук, доцент кафедры высшей математики
О.В. ТИТОВ[2], кандидат физико-математических наук, доцент кафедры естественных наук

[1]*УО «Белорусский торгово-экономический университет потребительской кооперации» г. Гомель, Республика Беларусь*
[2]*УО «Гомельский инженерный институт» МЧС Республики Беларусь, г. Гомель, Республика Беларусь*

Представлен способ выбора лучшего ресурса для хранения информации методом Мамдани.

**Ключевые слова:** нечеткая логика, метод Мамдани, оценки альтернатив, нечеткое регулирование, надежность.

В современном мире одной из важных задач является задача передачи и хранения информации.

Данная статья посвящается поиску оптимального ресурса для хранения информации. Роль и важность системы хранения определяются постоянно возрастающей ценностью информации в современном обществе. Возможность доступа к данным и управления ими является необходимым условием для выполнения любых задач. Безвозвратная потеря данных может привести к фатальным последствиям. Утраченные вычислительные ресурсы можно восстановить, а утраченные данные, при отсутствии грамотно спроектированной и внедренной системы резервирования, уже не подлежат восстановлению.

Разработка разного рода «распределенных файловых систем» в настоящее время является одной из активно развиваемых областей информатики. Большинство таких систем работает на основе изготовления множества полных копий данных, хранимых в разных местах, и обеспечения различных механизмов синхронизации этих данных.

Авторы статьи предлагают использовать аппарат нечеткой логики при выборе ресурса для хранения информации. Предложенный способ позволяет быстро и эффективно выбирать такой ресурс.

Практический опыт разработки систем нечеткого логического вывода свидетельствует, что сроки и стоимость их проектирования значительно меньше, чем при использовании традиционного математического аппарата. При этом обеспечивается требуемый уровень робастности и прозрачности моделей.

Каждый ресурс обладает следующими характеристиками: скорость доступа (будем измерять в Mb/s), надежность (процент времени непрерывной работы), концентрация (процент информации уже находящейся на ресурсе). Использование одного ресурса или сервиса для хранения всей информации авторы считают не целесообразным.

Методом Мамдани для каждого ресурса вычислим вероятность того, что на данный ресурс будет загружена очередная порция информации.

Для решения поставленной задачи введем следующие имена переменных: скорость, надежность, концентрация и вероятность. Определим нечеткие переменные и их функции принадлежности:

скорость высокая (рис. 1а), скорость не высокая (рис. 1б),

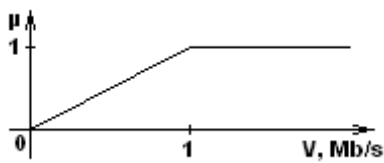 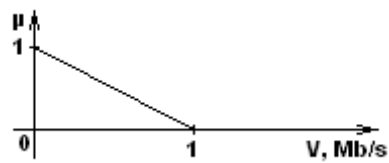

Рис. 1а                                        Рис. 1б

надежность высокая (рис. 2а), надежность не высокая (рис. 2б),

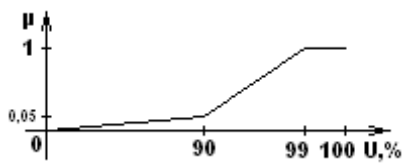 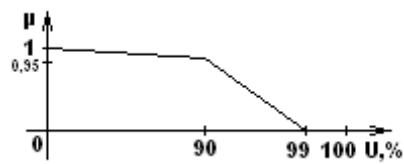

Рис. 2а                                        Рис. 2б

концентрация низкая (рис. 3а), концентрация не низкая (рис. 3б),

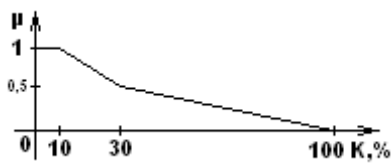 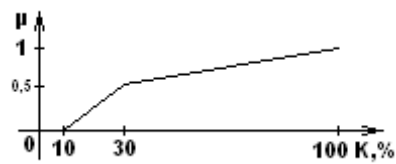

Рис. 3а                                        Рис. 3б

вероятность низкая (рис. 4а), вероятность высокая (рис. 4б).

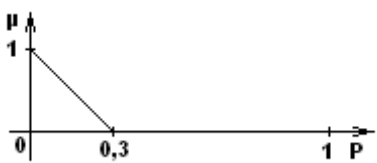 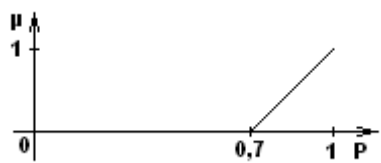

Рис. 4а                                        Рис. 4б

Графики функций принадлежности, приведенных выше нечетких переменных, основываются на экспертных оценках и аналитических выводах.

Введем два правила нечетких продукций:

1) ЕСЛИ (Скорость высокая) И (Надежность высокая) И (Концентрация низкая) ТО (Вероятность высокая).

2) ЕСЛИ (Скорость не высокая) И (Надежность не высокая) И (Концентрация не низкая) ТО (Вероятность низкая).

Выбранные правила нечетких продукций позволяют повышать (понижать) вероятность выбора ресурса, в зависимости от его актуальных характеристик.

Используя метод Мамдани для выше перечисленных нечетких переменных и правил нечетких продукций, мы получим для каждого i-го ресурса его вероятность, которую обозначим $p_i$. Для дефаззификация используется метод центра тяжести.

Далее можно поступить двумя способами:

1) выбрать ресурс с максимальной вероятностью, и записать туда следующую часть информации;

2) провести испытание для вероятностного пространства $\{\omega_i \mid P(\omega_i)=p_i/S\}$, где $S=\Sigma p_i$ и таким образом выбрать ресурс.

В данной статье приведены результаты анализа нечетких переменных, выраженных в графиках функций принадлежности для каждой из них.

Применение предложенного метода позволяет быстро и достаточно точно оценить надежность ресурса (сервиса) для хранения информации.

## Литература

**E.V. Legchekova, O.V. Titov**
**CHOOSING THE BEST RESOURCE BY METHOD OF MAMDANI**

A method for selecting the best service for the storage of information by Mamdani.